\documentclass[aps,prl,twocolumn,showpacs,amsmath,amssymb,superscriptaddress]{revtex4-1}

\usepackage{amsmath,amsfonts,bm,graphicx,verbatim,mathrsfs,units}
\usepackage[colorlinks=true,citecolor=blue]{hyperref}
\usepackage{pdfpages}

\newcommand{\ket}[1]{\left| #1 \right \rangle}
\newcommand{\bra}[1]{\left \langle #1 \right|}

\newcommand{\revision}[1]{#1}

\begin{document}
\title{Dynamical generation and detection of entanglement in neutral leviton pairs}
\date{\today}

\author{David Dasenbrook}
\affiliation{D\'epartement de Physique Th\'eorique, Universit\'e de Gen\`eve, 1211 Gen\`eve, Switzerland}
\author{Christian Flindt}
\affiliation{Department of Applied Physics, Aalto University, 00076 Aalto, Finland}

\begin{abstract}
  The entanglement of coherently split electron-hole pairs in an electronic conductor is typically
  not considered accessible due to particle number conservation and fermionic superselection rules.
  We demonstrate here that current cross-correlation measurements at the outputs of an electronic
  Mach-Zehnder interferometer can nevertheless provide a robust witness of electron-hole
  entanglement. Specifically, we consider neutral excitations generated by modulating the
  transmission of an unbiased quantum point contact periodically in time. For an optimized
  modulation profile, an entangled state with one positively-charged leviton (a hole) and one
  negatively-charged leviton (an electron) gets delocalized over the two paths of the interferometer
  and is detected at the output arms. We evaluate the influence of finite electronic temperatures
  and dephasing corresponding to recent experiments.
\end{abstract}



\maketitle

\paragraph{Introduction.---}

The development of dynamic single-electron emitters has triggered great interest in gigahertz
quantum electronics~\cite{bocquillon14}. Carefully engineered excitations can now be emitted on top
of the Fermi sea in an electronic conductor using driven mesoscopic capacitors
\cite{gabelli06,feve07,bocquillon13} or designed voltage pulses applied to an electrical
contact~\cite{dubois13nature,jullien14}. Lorentzian-shaped voltage pulses (or a linear drive of a
capacitor \cite{buttiker93,keeling08}) excite noiseless quasiparticles known as levitons without
accompanying electron-hole pairs~\cite{ivanov97,levitov96,keeling06,dubois13}. These experimental
advances open up intriguing avenues for quantum information processing with coherent electrons.  One
important goal is to generate and detect entanglement of levitons, borrowing ideas and concepts from
quantum optics.

Entangled modes with different photon numbers can be generated by sending single photons onto a beam
splitter~\cite{bjork01,vanenk05}. Due to particle number superselection rules this type of
entanglement has often been considered inaccessible~\cite{wiseman03}. Further investigations have
however clarified that entangled states of different photon numbers provide a resource that is as
useful as polarization-entangled photons \cite{lombardi02,bartlett07,salart10,takeda15}. Advancing
similar techniques to entangle  states of different electron numbers is clearly desirable, however,
the task is challenging. For example, a suitable witness to detect the electronic entanglement must
be identified. As such, earlier proposals have instead focused on the orbital entanglement of
several electron-hole pairs~\cite{beenakker03,samuelsson04,samuelsson05,sherkunov12} or pairs of
electrons~\cite{samuelsson03,sim06,samuelsson09}.

In this Rapid Communication we present an experimental recipe for the detection of electron-hole
entanglement in an electronic conductor. Specifically, we demonstrate that noise measurements at the
outputs of an electronic Mach-Zehnder interferometer, despite the fermionic superselection rules,
can provide a robust witness of electron-hole entanglement. An entangled state with one
positively-charged leviton and one negatively-charged leviton is produced at a quantum point contact
(QPC) and is delocalized across the two arms of the interferometer. Due to particle number
conservation the electron-hole entanglement in the state cannot easily be used to violate a Bell
inequality~\cite{brunner14}. We circumvent this problem by recombining the state at a second QPC. As
we show, an entanglement witness can be constructed from  cross-correlation measurements at the
output arms. We evaluate the entanglement witness using Floquet theory and find that the
electron-hole entanglement can be detected for realistic system parameters, including finite
electronic temperatures and dephasing corresponding to recent experiments
\cite{ji03,neder07,roulleau07,litvin08,huynh12}.

\begin{figure}
  \centering
  \includegraphics[width=0.88\columnwidth]{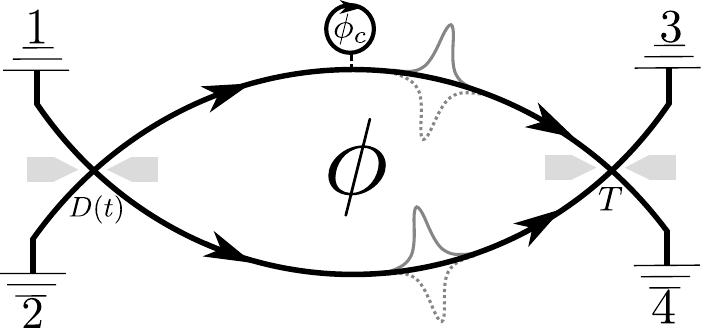}
  \caption{Dynamic Mach-Zehnder interferometer. Entangled states of neutral leviton-pairs are generated by modulating the transmission $D(t)=|d(t)|^2$ of the first QPC periodically in time. The levitons travel along edge states to the second QPC with transmission $T$ and the entanglement is detected by measuring the current cross-correlations at the output arms. The interferometer encloses the magnetic flux $\phi$. Negatively-charged levitons in the upper part can enter a small cavity with magnetic flux $\phi_c$ and pick up an additional phase.}
  \label{fig:setup}
\end{figure}

\paragraph{Mach-Zehnder interferometer.---}
The interferometer consists of a Corbino disk in the quantum Hall regime with electronic motion
along edge states from left to right in Fig.~\ref{fig:setup}. The upper and lower arms of the
interferometer form a loop that encloses the magnetic flux~$\phi$. In addition, electrons above the
Fermi level in the upper arm can enter a small cavity which encloses the flux $\phi_c$. Two QPCs act
as electronic beam splitters. Contrary to recent experiments, all contacts are grounded. Instead, we
modulate the transmission probability of the first QPC periodically in time in such a way that clean
electron-hole excitations are generated out of the otherwise undisturbed Fermi sea at the location
of the QPC. Each electron-hole pair delocalizes across the arms of the interferometer, leading to
a superposition of a negatively-charged leviton being in the upper arm and a positively-charged
leviton in the lower arm and vice versa.
As we go on to show, the resulting electron-hole entanglement can be detected by
measuring the cross-correlations of the currents in the output arms after the second QPC.

\paragraph{Dynamic entanglement generation.---}
We start with the generation of clean electron-hole pairs at a QPC. This problem is closely related to the creation of levitons by applying lorentzian-shaped voltage pulses to a contact \cite{ivanov97,levitov96,keeling06,dubois13}. As predicted by Levitov and co-workers and recently realized
experimentally~\cite{dubois13nature,jullien14}, pulses of the form
\begin{equation}
  \label{eq:Vt}
  V(t) = - \frac{\hbar}{e} \sum_{n=-\infty}^\infty \frac{2 \eta}{(t+n \mathcal{T})^2 + \eta^2},
\end{equation}
lead to the emission of levitons from the contact on top of the otherwise undisturbed Fermi sea (with a holelike leviton going into the contact). The width of the pulses is $\eta$ and $\mathcal{T}$ the period of the driving. Levitons are created as each electron leaving the contact picks up the phase factor $e^{i\varphi(t)}$ with the phase given as
\begin{equation}
  \label{eq:phase}
  \varphi(t) = - \frac{e}{\hbar} \int_0^t \mathrm{d} t' V(t').
\end{equation}
The phase changes sign $\varphi(t)\rightarrow -\varphi(t)$ upon inverting the voltage
$V(t)\rightarrow -V(t)$, leading to the emission of a holelike leviton from the contact.

Remarkably, a similar strategy can be used to generate superpositions of electronlike and holelike
levitons by modulating the transmission of a QPC periodically in time~\cite{sherkunov09,zhang09}. To
see this, we consider the time-dependent scattering matrix of the first QPC in Fig.~\ref{fig:setup},
\begin{equation}
  \label{eq:Stmatrix}
  S(t) = \left[\begin{array}{cc}
    r(t) & d(t) \\
    -d(t) & r(t)
  \end{array}\right],
\end{equation}
where the reflection and transmission amplitudes, chosen to be real
below, fulfill $|r(t)|^2+|d(t)|^2=1$. Switching to the eigenbasis of $S(t)$, particles in the two incoming eigenchannels will be
completely reflected with the reflection amplitudes $r(t) \pm i d(t)$ given by the eigenvalues
of $S(t)$. We now choose the transmission and reflection as
\begin{equation}
  \label{eq:st}
  \begin{split}
  d(t)&=\sin\varphi(t),\\
    r(t)&=\cos\varphi(t)
  \end{split}
\end{equation}
with $\varphi(t)$ given by Eq.~(\ref{eq:phase}). The reflection amplitudes in the eigenbasis then
become $r(t) \pm i d(t)=e^{\pm i\varphi(t)}$, implying that an electronlike leviton is reflected in
one eigenchannel and a holelike leviton in the other. Returning to the physical channels of the
QPC, the outgoing state after the small cavity in Fig.~\ref{fig:setup} becomes
\begin{equation}
  \label{eq:outgoingstate}
    \begin{split}
  \ket{\Psi} = &\frac{1}{2} \Big( e^{i \vartheta} \hat{b}_{u-}^\dagger \hat{b}_{u+}^\dagger -
    \hat{b}_{l-}^\dagger \hat{b}_{l+}^\dagger \\
    &+i \Big\{ e^{i \vartheta} \hat{b}_{u-}^\dagger
      \hat{b}_{l+}^\dagger + \hat{b}_{u+}^\dagger \hat{b}_{l-}^\dagger \Big\}\Big) \ket{0},
\end{split}
\end{equation}
where $\ket{0}$ is the filled Fermi sea at zero temperature and the Fermi energy is zero. The
operators $\hat{b}^\dagger_{i-} = \sum_{E>0} e^{-\eta E} \hat{b}_i^{\dagger}(E)$ and
$\hat{b}^\dagger_{i+} = \sum_{E<0} e^{\eta E} \hat{b}_i(E)$ create electronlike and holelike
levitons in the upper $(i=u)$ or lower $(i=l)$ arms of the interferometer, and
$\hat{b}_i^{\dagger}(E)$ creates electrons at energy $E$ in either of the two arms. We assume for
now that the effect of the small cavity can be encoded in a tuneable phase
$\vartheta=\vartheta(\phi_c)$ picked up by electronlike levitons in the upper arm. Below, we return
to a more detailed description of the cavity [see Eq.~(\ref{eq:cavity})]. {\revision As }
Eq.~\eqref{eq:outgoingstate} {\revision cannot be written as a product of $b_u^\dagger$ and
$b_l^\dagger$ operators, the state} is entangled. Furthermore, the projection of the state on the
subspace with one particle per arm (in curly brackets) is maximally entangled in the electron-hole
degree of freedom. This is the entanglement we wish to detect.  We note that the state in
Eq.~\eqref{eq:outgoingstate}  can also be generated by emitting levitons from different inputs onto
the QPC tuned to half transmission using quantum capacitors with a linear
drive~\cite{buttiker93,keeling08}.

\begin{figure*}
  \centering
  \includegraphics[width=.44\textwidth]{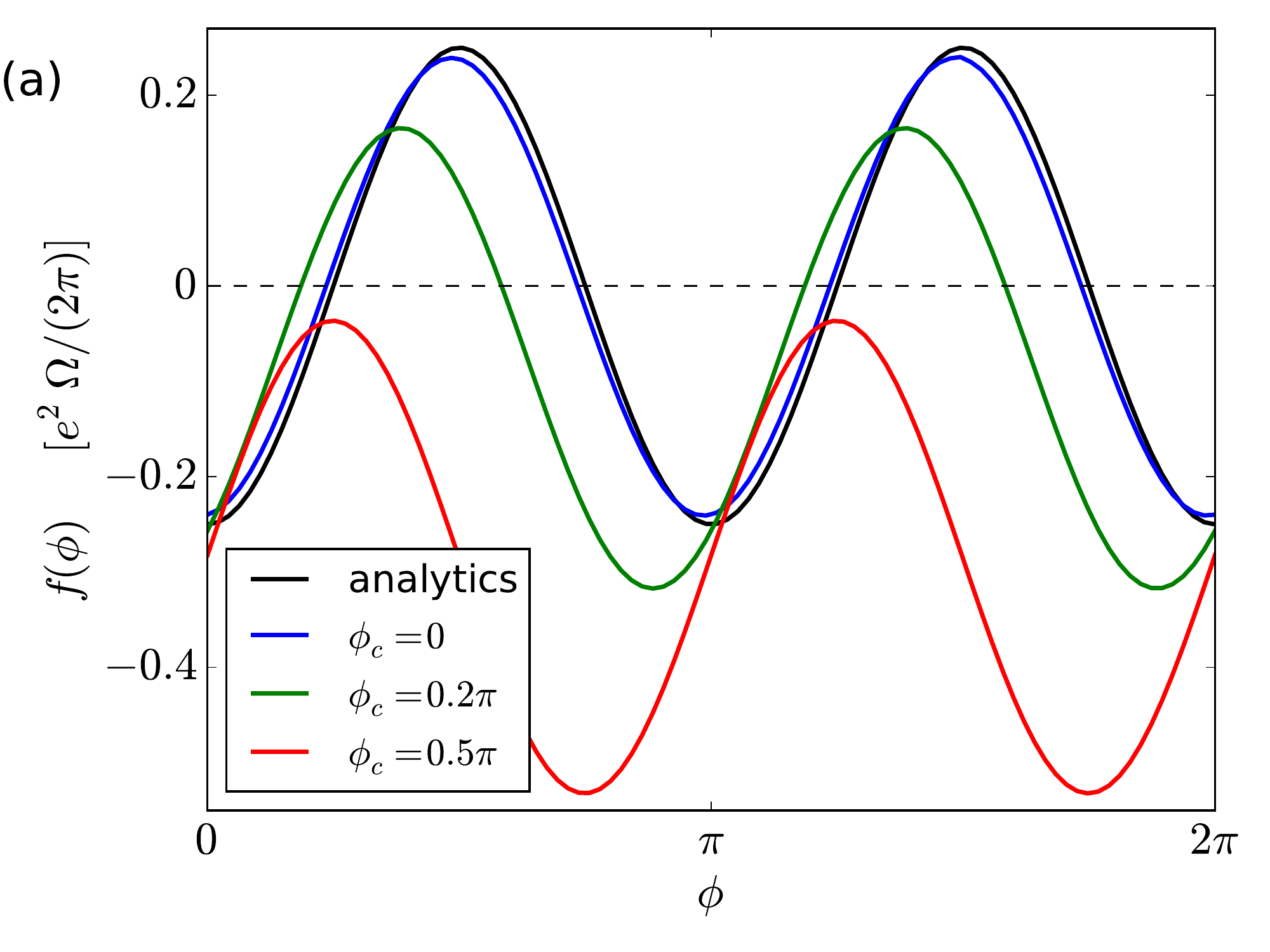}
  \includegraphics[width=.44\textwidth]{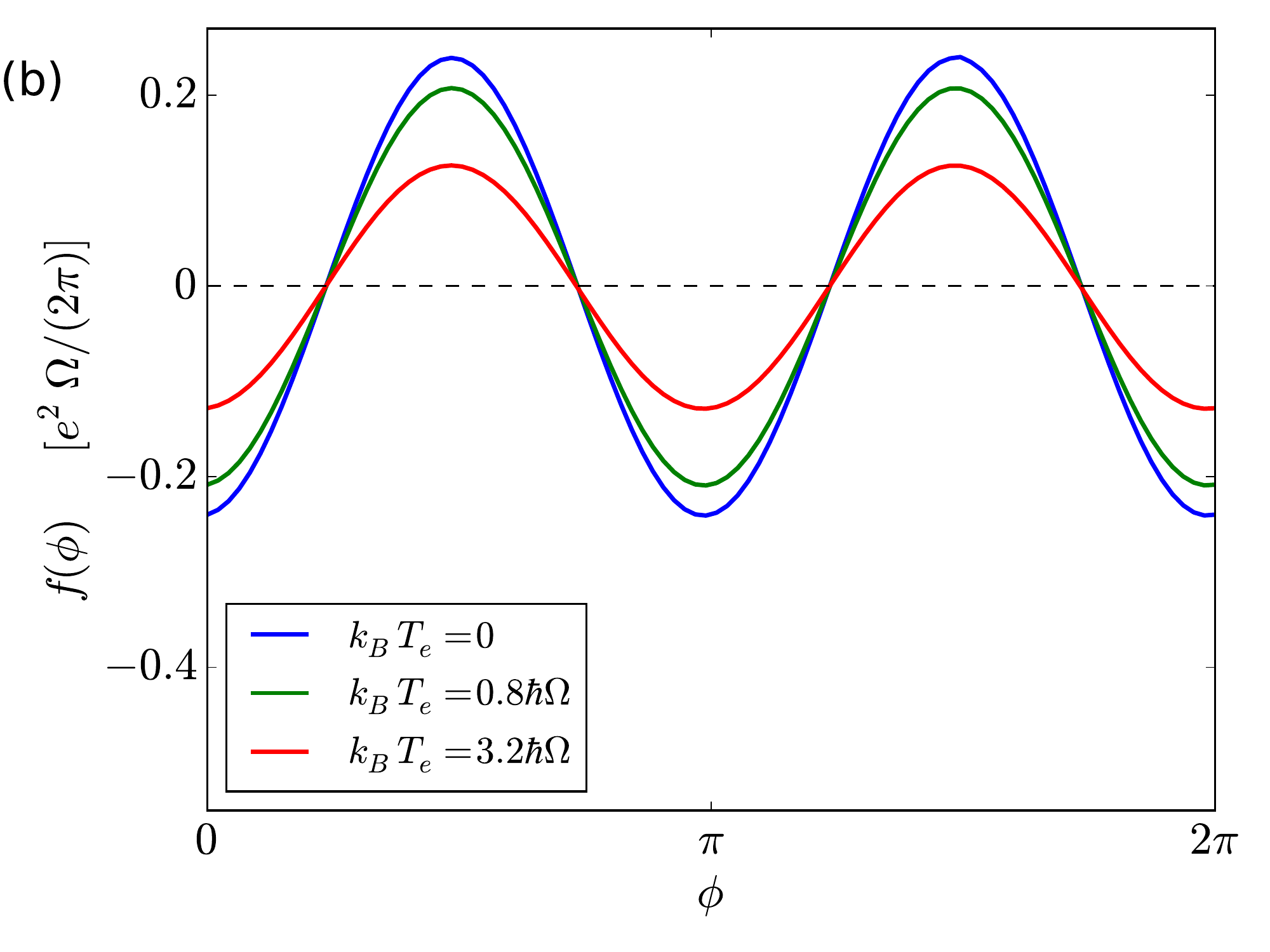}
  \caption{(color online). Entanglement witness. (a) The witness $f(\phi)$ as a function of the
  Aharonov-Bohm phase $\phi$ with different values of the flux $\phi_c$ enclosed by the small
  cavity. The electronic temperature is zero. Positive values of $f$ (above the dashed line) signal
  electron-hole entanglement. With $\phi_c=0$, the Floquet calculation is close to the analytic
  result -$\cos(2 \phi)/4$ (black line) corresponding to the maximally entangled state in
  Eq.~(\ref{eq:outgoingstate}). The parameters of the cavity are $\tau = \mathcal{T}/50$ and
  $\mathcal{B}=5\mathcal{T}/\hbar$. (b) The witness $f(\phi)$ as a function of $\phi$ with
  $\phi_c=0$ and different electronic temperatures $T_e$. An increased temperature merely decreases
  the amplitude of the oscillations, so that the entanglement is still detectable at finite
  temperatures.}
\label{fig:witness}
\end{figure*}

\paragraph{Entanglement witness.---}
It is difficult to formulate a Bell inequality for the entanglement between modes of different
particle numbers. (It would require measurements in a basis of nondefinite particle number and for
electrons a superconductor, for instance, would be needed \cite{beenakker14}). Furthermore, the
violation of Bell inequalities often relies on very high visibilities and is therefore currently out
of reach for mesoscopic conductors. To circumvent these problems, we instead construct an
entanglement witness based on current cross-correlation measurements at the output arms after the
second QPC, similar to what has  been considered in the context of spin entanglement
\cite{burkard03,giovanetti06}. To develop our witness, we consider a general two-leviton state
incident on the second QPC
\begin{equation}
  \label{eq:generalinputstate}
  \ket{\Upsilon} = \sum_{\stackrel{{\alpha,\beta=\pm}}{i,j=u,l}} \Upsilon_{\alpha \beta}^{ij} \hat{b}_{i
    \alpha}^\dagger \hat{b}_{j\beta}^\dagger \ket{0}
\end{equation}
with the normalization condition $\sum_{\alpha \beta i j} |\Upsilon_{\alpha \beta}^{ij}|^2 = 1$.
Equation~\eqref{eq:outgoingstate} is a particular example of such a state.

If the projection on the single-particle sector is separable, the matrix $\Upsilon^{ul}$ has rank
one~\cite{amico08}. Calculating the cross-correlator $S_{34}(\phi) = \langle \hat{I}_3 \hat{I}_4
\rangle$ measured after the second QPC, we can then show  that the function \footnote{See
supplemental material at \dots}
\begin{equation}
  \label{eq:fphi}
  f(\phi)\equiv S_{34}(\phi) - S_0(1-2TR)
\end{equation}

is always zero or negative. Here, $S_0 = \sum_{\alpha \beta} \alpha \beta |\Upsilon_{\alpha
\beta}^{ul}|^2$ is the noise at zero transmission and $\phi = 2 \pi \Phi/\Phi_0$ is the
Aharonov-Bohm phase with $\Phi$ being the magnetic flux enclosed by the interferometer and
$\Phi_0=h/e$ the magnetic flux quantum. Moreover, for a general separable density matrix $\hat{\rho}  =
\sum_n p_n \ket{\Upsilon_n}\! \bra{\Upsilon_n}$ \cite{horodecki09} with each $\ket{\Upsilon_n}$ of
the form~(\ref{eq:generalinputstate}) and separable, the noise is the average noise of each
separable state weighted by the probabilities $p_n$. Therefore, the condition $f(\phi)>0$ provides a
witness of electron-hole entanglement also at finite temperatures. The witness is not optimal, since
negative two-particle contributions to the noise [terms with $i=j=u$ or $i=j=l$ in
Eq.~\eqref{eq:generalinputstate}] can make it harder to detect the entanglement, even if it is
maximal~\cite{horodecki09}.  Importantly, our witness relies on reconnecting the two arms at the
second QPC, making the measurement nonlocal. This is the key ingredient that allows us to
circumvent the superselection rules for particle number.

Evaluating the witness for the state in Eq.~(\ref{eq:outgoingstate}), we first find a pure
interference current determined by the enclosed fluxes, $\langle \hat{I}_3 \rangle = - \langle
\hat{I}_4 \rangle = e(\Omega/\pi) \sqrt{TR} \cos[\vartheta(\phi_c)/2]
\sin[\phi+\vartheta(\phi_c)/2]$, where $\Omega=2\pi/\mathcal{T}$ is the frequency of the driving.
For the current cross-correlator, we find
\begin{equation}
  \label{eq:crossnoise}
  S_{34} = - \frac{e^2 \Omega}{4 \pi} [1-2TR\{1-\cos(2\phi)\}]+\langle \hat{I}_3 \rangle\langle \hat{I}_4 \rangle.
\end{equation}
Both the current and the noise are independent of the pulse width $\eta$, which determines the
spatial extent of the levitons. Now, tuning the phase $\vartheta(\phi_c)$ to~$\pi$ and choosing
$T=R=1/2$, the average currents vanish and the witness becomes $f(\phi) = -e^2\Omega/(8\pi)\cos(2
\phi)$, which clearly can be positive, signaling electron-hole entanglement.

\paragraph{Floquet scattering theory.---}
The noise in Eq.~\eqref{eq:crossnoise} corresponds to the ideal case of the entangled state in
Eq.~(\ref{eq:outgoingstate}). We now proceed with a full Floquet calculation
\cite{moskalets02,moskaletsbook} of the entanglement witness for the interferometer in
Fig.~\ref{fig:setup}, including finite electronic temperatures and a detailed description of the
small cavity. In this case, the outgoing state is not known and we need to evaluate the witness to
detect the entanglement. As we will see, electron-hole entanglement is detectable under realistic
experimental conditions.

The current operator in contact $i$ can be written as $\hat{I}_i = \frac{e}{h}\int_{-\infty}^\infty \mathrm{d} E\left(\hat{c}_i^\dagger(E)\hat{c}_i(E) -
  \hat{a}_i^\dagger(E) \hat{a}_i(E) \right)$
in terms of the operators for particles at energy $E$ incoming from and outgoing to reservoir $i$, respectively. The operators for outgoing particles can be expressed as
\begin{equation}
  \label{eq:outgoingincoming}
  \hat{c}_i(E) = \sum_j \sum_{n=-\infty}^\infty \mathcal{S}_{ij}(E,E_n) \hat{a}_j(E_n),
\end{equation}
where $\mathcal{S}$ is the Floquet scattering matrix and $\hat{a}_j(E)$ are operators for incoming particles from reservoir $j$.

The Floquet amplitudes for incoming particles at energy $E$ to scatter into the outgoing reservoirs with energy $E_n = E + n \hbar \Omega$, having absorbed ($n>0$) or emitted ($n<0$) $|n|$ energy quanta of size $\hbar\Omega$, read
\begin{equation}
  \label{eq:floquetsmatrix}
  \begin{split}
  \mathcal{S}_{31}(E_n,E) &= -\sqrt{T} S_F^d(n) + \sqrt{R} e^{i \phi} S_c(E_n) S_F^r(n), \\
  \mathcal{S}_{32}(E_n,E) &= \sqrt{T} S_F^r(n) - \sqrt{R} e^{i \phi} S_c(E_n) S_F^d(n), \\
  \mathcal{S}_{41}(E_n,E) &= \sqrt{R} S_F^d(n) + \sqrt{T} e^{i \phi} S_c(E_n) S_F^r(n), \\
  \mathcal{S}_{42}(E_n,E) &= -\sqrt{R} S_F^r(n) - \sqrt{T} e^{i \phi} S_c(E_n) S_F^d(n).
  \end{split}
\end{equation}
Here, the Floquet amplitudes of the first QPC, $S_F^{s}(n) = \int_{0}^\mathcal{T} \mathrm{d} t s(t) e^{i n \Omega t} / \mathcal{T}$, with $s=d,r$ given in Eq.~(\ref{eq:st}), are
$S_F^d(n\neq 0) = -\sinh(2\pi\eta) e^{-2 \pi |n| \eta}$, $S_F^d(n=0) = e^{-2 \pi \eta}$ and $S_F^r(n) = \mathrm{sgn}(n)\sinh(2\pi\eta) e^{-2 \pi |n| \eta}$. In addition, the scattering matrix of the small cavity reads
\begin{equation}
\label{eq:cavity}
S_c(E)=r_c(E)+t_c^2(E)\frac{e^{i(\phi_c+E\tau/\hbar+\pi)}}{1+r_c(E)e^{i(\phi_c+ E\tau/\hbar+\pi)}},
\end{equation}
where $\tau$ is the time it takes to complete one loop inside the cavity and $t_c(E) = 1/[\exp(-\mathcal{B} E)+1]$ is the transmission amplitude into the cavity with the cut-off $\mathcal{B}$ being tunable by a magnetic field \cite{fertig87,buttiker90}. The reflection amplitude is $r_c(E)=\sqrt{1-t_c^2(E)}$. With a sharp cut-off $\mathcal{B}\gg \mathcal{T}/h$ and a short loop-time $\tau\ll \mathcal{T}$, we recover the state in Eq.~(\ref{eq:outgoingstate}) with $\vartheta(\phi_c)\simeq \phi_c+\pi$.

Figure~\ref{fig:witness} shows the entanglement witness calculated using Floquet scattering
theory~\cite{moskalets02,moskaletsbook}. We vary the Aharonov-Bohm phase $\phi$ and show in panel
(a) results for different values of the flux $\phi_c$ enclosed by the cavity. The electronic
temperature is zero. The entanglement cannot be detected in all situations. However, by tuning
$\phi_c$ we come close to the analytic result (black line) corresponding to the maximally entangled
state in Eq.~(\ref{eq:outgoingstate}). The system then maximally violates the inequality
$f(\phi)\leq 0$ in the sense that the witness has the same weight above and below the $f=0$ line as
a function of~$\phi$. Under this condition, the witness is thus expected to be very
robust against a decreased visibility, in contrast to entanglement detection based on Bell
inequalities \cite{samuelsson09}.

In panel (b) we fix the optimal value of $\phi_c=0$ and consider the effect of a finite electronic
temperature. With increasing temperature, the amplitude of the oscillations decreases and the
entanglement gets harder to detect. Still, even with temperatures that are higher than the driving
frequency, the witness can become positive and entanglement can be detected. Since there is no
direct scattering path between the two output reservoirs, thermal noise is not visible in the
witness \cite{hofer14}. The results in Fig.~\ref{fig:witness} are promising for the detection of
electron-hole entanglement in driven mesoscopic conductors.

\begin{figure}
  \centering
  \includegraphics[width=\columnwidth]{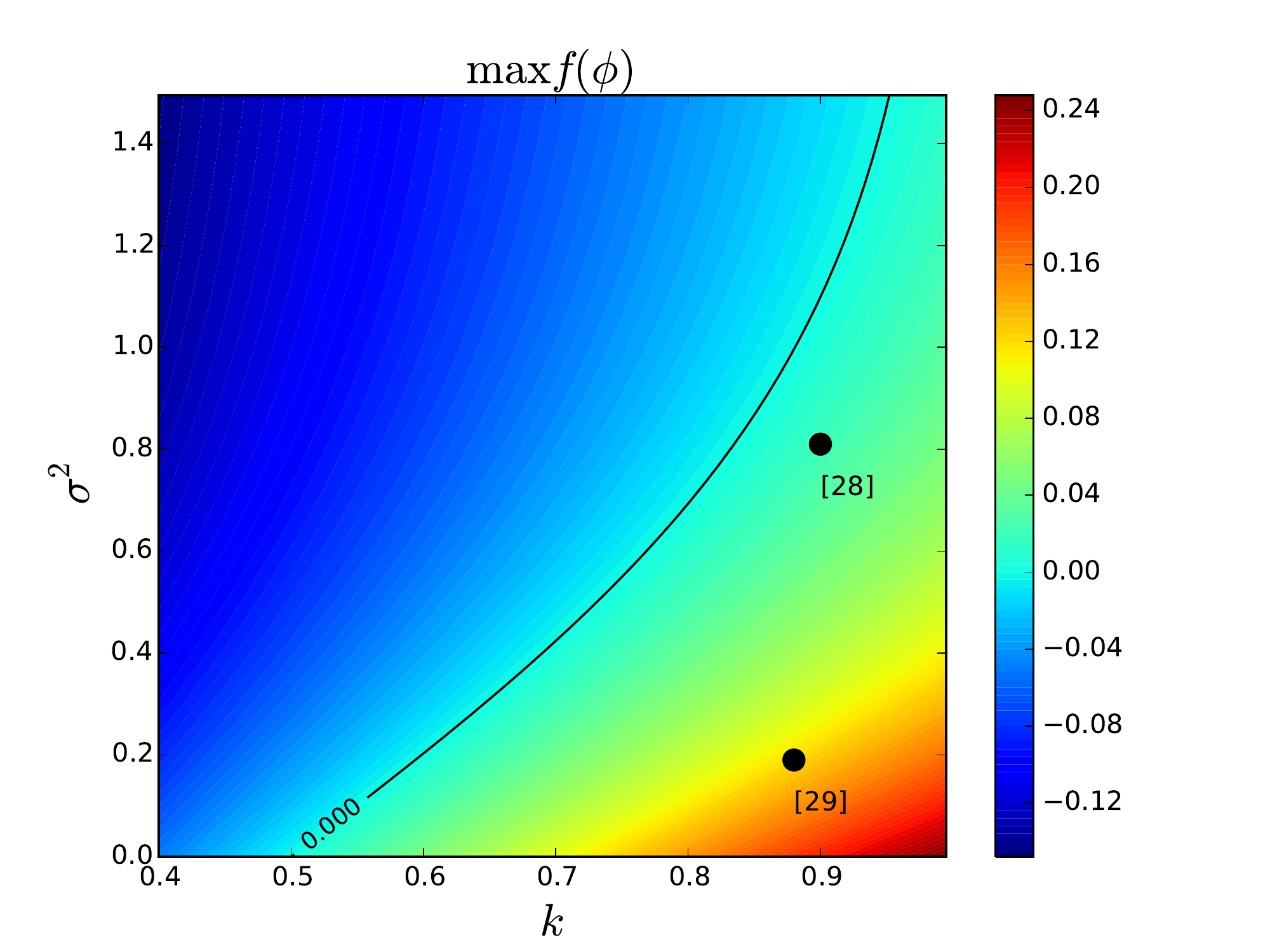}
  \caption{(color online). Influence of dephasing. The maximum value of the entanglement witness $f(\phi)$ as a function of the decoherence parameter $k$ and the variance of the phase~$\sigma^2$. The contour line separates the region of detectable electron-hole entanglement, where the witness is positive, from the region where the entanglement cannot be detected. The black dots mark the experimental parameters from Refs.~\cite{ji03,neder07}, which lie in the region of detectable entanglement.}
  \label{fig:dephasing}
\end{figure}

\paragraph{Dephasing mechanisms---}
Finally, to estimate the influence of dephasing and phase averaging, we return to the analytic
result for the noise in Eq.~\eqref{eq:crossnoise}. Focusing on the optimal value $\vartheta=\pi$,
the noise can be written as $S_{34} = -e^2\Omega/(4\pi)[1- 2kTR \{1 - \exp(-2 \sigma^2) \cos(2\phi) \}]$
in terms of the phenomenological parameters $k$ \cite{ji03} and $\sigma^2$~\cite{roulleau07} which
describe the coherence of the wave function across the interferometer ($k=1$ meaning full coherence
and $k=0$ no coherence, e.~g.~as a result of a finite electronic temperature or interactions) and
the variance of the total Aharonov-Bohm phase leading to phase-averaging. In
Fig.~\ref{fig:dephasing} we show the maximal value of the witness $f(\phi)$ as a function of $k$
and~$\sigma^2$. We find that the witness is robust against moderate dephasing mechanisms and that
entanglement is detectable for parameters corresponding to the experiments reported in
Refs.~\cite{ji03,neder07}. {\revision This is in contrast to the detection of orbital entanglement based on a Bell inequality \cite{samuelsson09}.}

\paragraph{Conclusions.---}
We have proposed and analyzed a dynamical scheme to generate and detect entanglement in the
electron-hole degree of freedom of leviton-pairs. Measuring the cross-correlations at the output
arms of a mesoscopic Mach-Zehnder interferometer, entanglement can be detected despite
superselection rules. The entanglement witness is robust against moderate dephasing mechanisms and
entanglement can be detected using current technologies. Future work may investigate the
entanglement entropy generated in this scheme~\cite{klich09,song11,song12,thomas15}.

\paragraph{Acknowledgments.---}
\begin{acknowledgments}
  We thank J.~Bowles, J.~B.~Brask, N.~Brunner, D.~C.~Glattli, P.~P.~Hofer, J.~R.~Ott, P.~Roulleau, and K.~H.~Thomas for helpful discussions. CF is affiliated with Centre for Quantum Engineering at Aalto University. DD gratefully acknowledges the hospitality of Aalto University. This work was supported by the Swiss NSF and Academy of Finland.
\end{acknowledgments}

\includepdf[pages={{},1}]{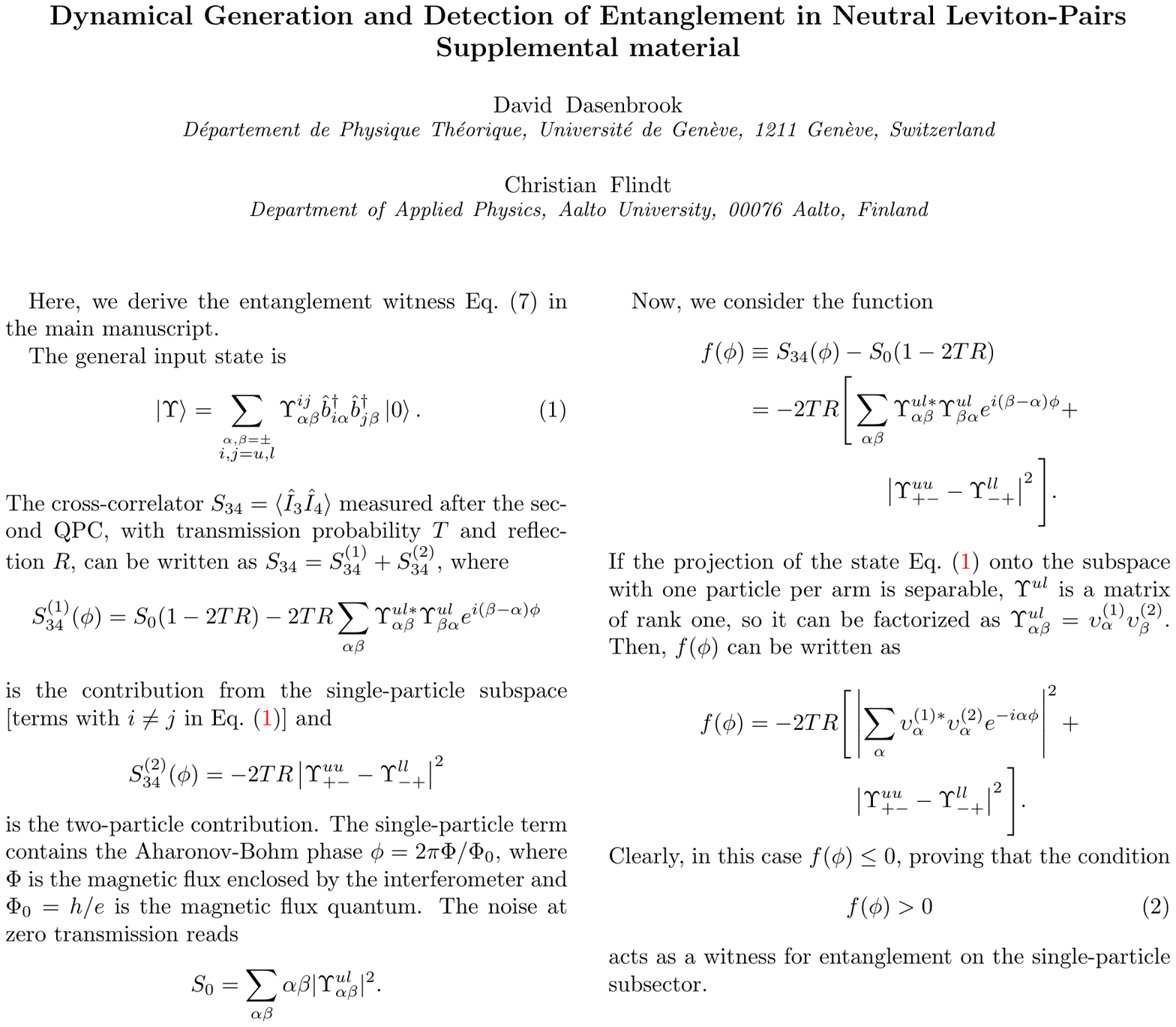}


\begin{thebibliography}{49}%
\makeatletter
\providecommand \@ifxundefined [1]{%
 \@ifx{#1\undefined}
}%
\providecommand \@ifnum [1]{%
 \ifnum #1\expandafter \@firstoftwo
 \else \expandafter \@secondoftwo
 \fi
}%
\providecommand \@ifx [1]{%
 \ifx #1\expandafter \@firstoftwo
 \else \expandafter \@secondoftwo
 \fi
}%
\providecommand \natexlab [1]{#1}%
\providecommand \enquote  [1]{``#1''}%
\providecommand \bibnamefont  [1]{#1}%
\providecommand \bibfnamefont [1]{#1}%
\providecommand \citenamefont [1]{#1}%
\providecommand \href@noop [0]{\@secondoftwo}%
\providecommand \href [0]{\begingroup \@sanitize@url \@href}%
\providecommand \@href[1]{\@@startlink{#1}\@@href}%
\providecommand \@@href[1]{\endgroup#1\@@endlink}%
\providecommand \@sanitize@url [0]{\catcode `\\12\catcode `\$12\catcode
  `\&12\catcode `\#12\catcode `\^12\catcode `\_12\catcode `\%12\relax}%
\providecommand \@@startlink[1]{}%
\providecommand \@@endlink[0]{}%
\providecommand \url  [0]{\begingroup\@sanitize@url \@url }%
\providecommand \@url [1]{\endgroup\@href {#1}{\urlprefix }}%
\providecommand \urlprefix  [0]{URL }%
\providecommand \Eprint [0]{\href }%
\providecommand \doibase [0]{http://dx.doi.org/}%
\providecommand \selectlanguage [0]{\@gobble}%
\providecommand \bibinfo  [0]{\@secondoftwo}%
\providecommand \bibfield  [0]{\@secondoftwo}%
\providecommand \translation [1]{[#1]}%
\providecommand \BibitemOpen [0]{}%
\providecommand \bibitemStop [0]{}%
\providecommand \bibitemNoStop [0]{.\EOS\space}%
\providecommand \EOS [0]{\spacefactor3000\relax}%
\providecommand \BibitemShut  [1]{\csname bibitem#1\endcsname}%
\let\auto@bib@innerbib\@empty
\bibitem [{\citenamefont {Bocquillon}\ \emph {et~al.}(2014)\citenamefont
  {Bocquillon}, \citenamefont {Freulon}, \citenamefont {Parmentier},
  \citenamefont {Berroir}, \citenamefont {Placais}, \citenamefont {Wahl},
  \citenamefont {Rech}, \citenamefont {Jonckheere}, \citenamefont {Martin},
  \citenamefont {Grenier}, \citenamefont {Ferraro}, \citenamefont
  {Degiovanni},\ and\ \citenamefont {F\`eve}}]{bocquillon14}%
  \BibitemOpen
  \bibfield  {author} {\bibinfo {author} {\bibfnamefont {E.}~\bibnamefont
  {Bocquillon}}, \bibinfo {author} {\bibfnamefont {V.}~\bibnamefont {Freulon}},
  \bibinfo {author} {\bibfnamefont {F.~D.}\ \bibnamefont {Parmentier}},
  \bibinfo {author} {\bibfnamefont {J.-M.}\ \bibnamefont {Berroir}}, \bibinfo
  {author} {\bibfnamefont {B.}~\bibnamefont {Placais}}, \bibinfo {author}
  {\bibfnamefont {C.}~\bibnamefont {Wahl}}, \bibinfo {author} {\bibfnamefont
  {J.}~\bibnamefont {Rech}}, \bibinfo {author} {\bibfnamefont {T.}~\bibnamefont
  {Jonckheere}}, \bibinfo {author} {\bibfnamefont {T.}~\bibnamefont {Martin}},
  \bibinfo {author} {\bibfnamefont {C.}~\bibnamefont {Grenier}}, \bibinfo
  {author} {\bibfnamefont {D.}~\bibnamefont {Ferraro}}, \bibinfo {author}
  {\bibfnamefont {P.}~\bibnamefont {Degiovanni}}, \ and\ \bibinfo {author}
  {\bibfnamefont {G.}~\bibnamefont {F\`eve}},\ }\href {\doibase
  10.1002/andp.201300181} {\bibfield  {journal} {\bibinfo  {journal} {Ann.
  Phys.}\ }\textbf {\bibinfo {volume} {526}},\ \bibinfo {pages} {1} (\bibinfo
  {year} {2014})}\BibitemShut {NoStop}%
\bibitem [{\citenamefont {Gabelli}\ \emph {et~al.}(2006)\citenamefont
  {Gabelli}, \citenamefont {F\`eve}, \citenamefont {Berroir}, \citenamefont
  {Pla{\c{c}}ais}, \citenamefont {Cavanna}, \citenamefont {Etienne},
  \citenamefont {Jin},\ and\ \citenamefont {Glattli}}]{gabelli06}%
  \BibitemOpen
  \bibfield  {author} {\bibinfo {author} {\bibfnamefont {J.}~\bibnamefont
  {Gabelli}}, \bibinfo {author} {\bibfnamefont {G.}~\bibnamefont {F\`eve}},
  \bibinfo {author} {\bibfnamefont {J.-M.}\ \bibnamefont {Berroir}}, \bibinfo
  {author} {\bibfnamefont {B.}~\bibnamefont {Pla{\c{c}}ais}}, \bibinfo {author}
  {\bibfnamefont {A.}~\bibnamefont {Cavanna}}, \bibinfo {author} {\bibfnamefont
  {B.}~\bibnamefont {Etienne}}, \bibinfo {author} {\bibfnamefont
  {Y.}~\bibnamefont {Jin}}, \ and\ \bibinfo {author} {\bibfnamefont {D.~C.}\
  \bibnamefont {Glattli}},\ }\href {\doibase 10.1126/science.1126940}
  {\bibfield  {journal} {\bibinfo  {journal} {Science}\ }\textbf {\bibinfo
  {volume} {313}},\ \bibinfo {pages} {499} (\bibinfo {year}
  {2006})}\BibitemShut {NoStop}%
\bibitem [{\citenamefont {F\`eve}\ \emph {et~al.}(2007)\citenamefont {F\`eve},
  \citenamefont {Mah\`e}, \citenamefont {Berroir}, \citenamefont {Kontos},
  \citenamefont {Pla\c{c}ais}, \citenamefont {Glattli}, \citenamefont
  {Cavanna}, \citenamefont {Etienne},\ and\ \citenamefont {Jin}}]{feve07}%
  \BibitemOpen
  \bibfield  {author} {\bibinfo {author} {\bibfnamefont {G.}~\bibnamefont
  {F\`eve}}, \bibinfo {author} {\bibfnamefont {A.}~\bibnamefont {Mah\`e}},
  \bibinfo {author} {\bibfnamefont {J.-M.}\ \bibnamefont {Berroir}}, \bibinfo
  {author} {\bibfnamefont {T.}~\bibnamefont {Kontos}}, \bibinfo {author}
  {\bibfnamefont {B.}~\bibnamefont {Pla\c{c}ais}}, \bibinfo {author}
  {\bibfnamefont {D.~C.}\ \bibnamefont {Glattli}}, \bibinfo {author}
  {\bibfnamefont {A.}~\bibnamefont {Cavanna}}, \bibinfo {author} {\bibfnamefont
  {B.}~\bibnamefont {Etienne}}, \ and\ \bibinfo {author} {\bibfnamefont
  {Y.}~\bibnamefont {Jin}},\ }\href {\doibase 10.1126/science.1141243}
  {\bibfield  {journal} {\bibinfo  {journal} {Science}\ }\textbf {\bibinfo
  {volume} {316}},\ \bibinfo {pages} {1169} (\bibinfo {year}
  {2007})}\BibitemShut {NoStop}%
\bibitem [{\citenamefont {Bocquillon}\ \emph {et~al.}(2013)\citenamefont
  {Bocquillon}, \citenamefont {Freulon}, \citenamefont {Berroir}, \citenamefont
  {Degiovanni}, \citenamefont {Pla\c{c}ais}, \citenamefont {Cavanna},
  \citenamefont {Jin},\ and\ \citenamefont {F\`eve}}]{bocquillon13}%
  \BibitemOpen
  \bibfield  {author} {\bibinfo {author} {\bibfnamefont {E.}~\bibnamefont
  {Bocquillon}}, \bibinfo {author} {\bibfnamefont {V.}~\bibnamefont {Freulon}},
  \bibinfo {author} {\bibfnamefont {J.-M.}\ \bibnamefont {Berroir}}, \bibinfo
  {author} {\bibfnamefont {P.}~\bibnamefont {Degiovanni}}, \bibinfo {author}
  {\bibfnamefont {B.}~\bibnamefont {Pla\c{c}ais}}, \bibinfo {author}
  {\bibfnamefont {A.}~\bibnamefont {Cavanna}}, \bibinfo {author} {\bibfnamefont
  {Y.}~\bibnamefont {Jin}}, \ and\ \bibinfo {author} {\bibfnamefont
  {G.}~\bibnamefont {F\`eve}},\ }\href {\doibase 10.1126/science.1232572}
  {\bibfield  {journal} {\bibinfo  {journal} {Science}\ }\textbf {\bibinfo
  {volume} {339}},\ \bibinfo {pages} {1054} (\bibinfo {year}
  {2013})}\BibitemShut {NoStop}%
\bibitem [{\citenamefont {Dubois}\ \emph
  {et~al.}(2013{\natexlab{a}})\citenamefont {Dubois}, \citenamefont {Jullien},
  \citenamefont {Portier}, \citenamefont {Roche}, \citenamefont {Cavanna},
  \citenamefont {Jin}, \citenamefont {Wegscheider}, \citenamefont {Roulleau},\
  and\ \citenamefont {Glattli}}]{dubois13nature}%
  \BibitemOpen
  \bibfield  {author} {\bibinfo {author} {\bibfnamefont {J.}~\bibnamefont
  {Dubois}}, \bibinfo {author} {\bibfnamefont {T.}~\bibnamefont {Jullien}},
  \bibinfo {author} {\bibfnamefont {F.}~\bibnamefont {Portier}}, \bibinfo
  {author} {\bibfnamefont {P.}~\bibnamefont {Roche}}, \bibinfo {author}
  {\bibfnamefont {A.}~\bibnamefont {Cavanna}}, \bibinfo {author} {\bibfnamefont
  {Y.}~\bibnamefont {Jin}}, \bibinfo {author} {\bibfnamefont {W.}~\bibnamefont
  {Wegscheider}}, \bibinfo {author} {\bibfnamefont {P.}~\bibnamefont
  {Roulleau}}, \ and\ \bibinfo {author} {\bibfnamefont {D.~C.}\ \bibnamefont
  {Glattli}},\ }\href {\doibase 10.1038/nature12713} {\bibfield  {journal}
  {\bibinfo  {journal} {Nature}\ }\textbf {\bibinfo {volume} {502}},\ \bibinfo
  {pages} {659} (\bibinfo {year} {2013}{\natexlab{a}})}\BibitemShut {NoStop}%
\bibitem [{\citenamefont {Jullien}\ \emph {et~al.}(2014)\citenamefont
  {Jullien}, \citenamefont {Roulleau}, \citenamefont {Roche}, \citenamefont
  {Cavanna}, \citenamefont {Jin},\ and\ \citenamefont {Glattli}}]{jullien14}%
  \BibitemOpen
  \bibfield  {author} {\bibinfo {author} {\bibfnamefont {T.}~\bibnamefont
  {Jullien}}, \bibinfo {author} {\bibfnamefont {P.}~\bibnamefont {Roulleau}},
  \bibinfo {author} {\bibfnamefont {B.}~\bibnamefont {Roche}}, \bibinfo
  {author} {\bibfnamefont {A.}~\bibnamefont {Cavanna}}, \bibinfo {author}
  {\bibfnamefont {Y.}~\bibnamefont {Jin}}, \ and\ \bibinfo {author}
  {\bibfnamefont {D.~C.}\ \bibnamefont {Glattli}},\ }\href {\doibase
  10.1038/nature13821} {\bibfield  {journal} {\bibinfo  {journal} {Nature}\
  }\textbf {\bibinfo {volume} {514}},\ \bibinfo {pages} {603} (\bibinfo {year}
  {2014})}\BibitemShut {NoStop}%
\bibitem [{\citenamefont {B\"uttiker}\ \emph {et~al.}(1993)\citenamefont
  {B\"uttiker}, \citenamefont {Thomas},\ and\ \citenamefont
  {Pr\^etre}}]{buttiker93}%
  \BibitemOpen
  \bibfield  {author} {\bibinfo {author} {\bibfnamefont {M.}~\bibnamefont
  {B\"uttiker}}, \bibinfo {author} {\bibfnamefont {H.}~\bibnamefont {Thomas}},
  \ and\ \bibinfo {author} {\bibfnamefont {A.}~\bibnamefont {Pr\^etre}},\
  }\href {\doibase 10.1016/0375-9601(93)91193-9} {\bibfield  {journal}
  {\bibinfo  {journal} {Phys. Lett. A}\ }\textbf {\bibinfo {volume} {180}},\
  \bibinfo {pages} {364} (\bibinfo {year} {1993})}\BibitemShut {NoStop}%
\bibitem [{\citenamefont {Keeling}\ \emph {et~al.}(2008)\citenamefont
  {Keeling}, \citenamefont {Shytov},\ and\ \citenamefont
  {Levitov}}]{keeling08}%
  \BibitemOpen
  \bibfield  {author} {\bibinfo {author} {\bibfnamefont {J.}~\bibnamefont
  {Keeling}}, \bibinfo {author} {\bibfnamefont {A.~V.}\ \bibnamefont {Shytov}},
  \ and\ \bibinfo {author} {\bibfnamefont {L.~S.}\ \bibnamefont {Levitov}},\
  }\href {\doibase 10.1103/PhysRevLett.101.196404} {\bibfield  {journal}
  {\bibinfo  {journal} {Phys. Rev. Lett.}\ }\textbf {\bibinfo {volume} {101}},\
  \bibinfo {pages} {196404} (\bibinfo {year} {2008})}\BibitemShut {NoStop}%
\bibitem [{\citenamefont {Ivanov}\ \emph {et~al.}(1997)\citenamefont {Ivanov},
  \citenamefont {Lee},\ and\ \citenamefont {Levitov}}]{ivanov97}%
  \BibitemOpen
  \bibfield  {author} {\bibinfo {author} {\bibfnamefont {D.~A.}\ \bibnamefont
  {Ivanov}}, \bibinfo {author} {\bibfnamefont {H.~W.}\ \bibnamefont {Lee}}, \
  and\ \bibinfo {author} {\bibfnamefont {L.~S.}\ \bibnamefont {Levitov}},\
  }\href {\doibase 10.1103/PhysRevB.56.6839} {\bibfield  {journal} {\bibinfo
  {journal} {Phys. Rev. B}\ }\textbf {\bibinfo {volume} {56}},\ \bibinfo
  {pages} {6839} (\bibinfo {year} {1997})}\BibitemShut {NoStop}%
\bibitem [{\citenamefont {Levitov}\ \emph {et~al.}(1996)\citenamefont
  {Levitov}, \citenamefont {Lee},\ and\ \citenamefont {Lesovik}}]{levitov96}%
  \BibitemOpen
  \bibfield  {author} {\bibinfo {author} {\bibfnamefont {L.~S.}\ \bibnamefont
  {Levitov}}, \bibinfo {author} {\bibfnamefont {H.}~\bibnamefont {Lee}}, \ and\
  \bibinfo {author} {\bibfnamefont {G.~B.}\ \bibnamefont {Lesovik}},\ }\href
  {\doibase 10.1063/1.531672} {\bibfield  {journal} {\bibinfo  {journal} {J.
  Math. Phys.}\ }\textbf {\bibinfo {volume} {37}},\ \bibinfo {pages} {4845}
  (\bibinfo {year} {1996})}\BibitemShut {NoStop}%
\bibitem [{\citenamefont {Keeling}\ \emph {et~al.}(2006)\citenamefont
  {Keeling}, \citenamefont {Klich},\ and\ \citenamefont {Levitov}}]{keeling06}%
  \BibitemOpen
  \bibfield  {author} {\bibinfo {author} {\bibfnamefont {J.}~\bibnamefont
  {Keeling}}, \bibinfo {author} {\bibfnamefont {I.}~\bibnamefont {Klich}}, \
  and\ \bibinfo {author} {\bibfnamefont {L.~S.}\ \bibnamefont {Levitov}},\
  }\href {\doibase 10.1103/PhysRevLett.97.116403} {\bibfield  {journal}
  {\bibinfo  {journal} {Phys. Rev. Lett.}\ }\textbf {\bibinfo {volume} {97}},\
  \bibinfo {pages} {116403} (\bibinfo {year} {2006})}\BibitemShut {NoStop}%
\bibitem [{\citenamefont {Dubois}\ \emph
  {et~al.}(2013{\natexlab{b}})\citenamefont {Dubois}, \citenamefont {Jullien},
  \citenamefont {Grenier}, \citenamefont {Degiovanni}, \citenamefont
  {Roulleau},\ and\ \citenamefont {Glattli}}]{dubois13}%
  \BibitemOpen
  \bibfield  {author} {\bibinfo {author} {\bibfnamefont {J.}~\bibnamefont
  {Dubois}}, \bibinfo {author} {\bibfnamefont {T.}~\bibnamefont {Jullien}},
  \bibinfo {author} {\bibfnamefont {C.}~\bibnamefont {Grenier}}, \bibinfo
  {author} {\bibfnamefont {P.}~\bibnamefont {Degiovanni}}, \bibinfo {author}
  {\bibfnamefont {P.}~\bibnamefont {Roulleau}}, \ and\ \bibinfo {author}
  {\bibfnamefont {D.~C.}\ \bibnamefont {Glattli}},\ }\href {\doibase
  10.1103/PhysRevB.88.085301} {\bibfield  {journal} {\bibinfo  {journal} {Phys.
  Rev. B}\ }\textbf {\bibinfo {volume} {88}},\ \bibinfo {pages} {085301}
  (\bibinfo {year} {2013}{\natexlab{b}})}\BibitemShut {NoStop}%
\bibitem [{\citenamefont {Bj\"ork}\ \emph {et~al.}(2001)\citenamefont
  {Bj\"ork}, \citenamefont {Jonsson},\ and\ \citenamefont
  {S\'anchez-Soto}}]{bjork01}%
  \BibitemOpen
  \bibfield  {author} {\bibinfo {author} {\bibfnamefont {G.}~\bibnamefont
  {Bj\"ork}}, \bibinfo {author} {\bibfnamefont {P.}~\bibnamefont {Jonsson}}, \
  and\ \bibinfo {author} {\bibfnamefont {L.~L.}\ \bibnamefont
  {S\'anchez-Soto}},\ }\href {\doibase 10.1103/PhysRevA.64.042106} {\bibfield
  {journal} {\bibinfo  {journal} {Phys. Rev. A}\ }\textbf {\bibinfo {volume}
  {64}},\ \bibinfo {pages} {042106} (\bibinfo {year} {2001})}\BibitemShut
  {NoStop}%
\bibitem [{\citenamefont {van Enk}(2005)}]{vanenk05}%
  \BibitemOpen
  \bibfield  {author} {\bibinfo {author} {\bibfnamefont {S.~J.}\ \bibnamefont
  {van Enk}},\ }\href {\doibase 10.1103/PhysRevA.72.064306} {\bibfield
  {journal} {\bibinfo  {journal} {Phys. Rev. A}\ }\textbf {\bibinfo {volume}
  {72}},\ \bibinfo {pages} {064306} (\bibinfo {year} {2005})}\BibitemShut
  {NoStop}%
\bibitem [{\citenamefont {Wiseman}\ and\ \citenamefont
  {Vaccaro}(2003)}]{wiseman03}%
  \BibitemOpen
  \bibfield  {author} {\bibinfo {author} {\bibfnamefont {H.~M.}\ \bibnamefont
  {Wiseman}}\ and\ \bibinfo {author} {\bibfnamefont {J.~A.}\ \bibnamefont
  {Vaccaro}},\ }\href {\doibase 10.1103/PhysRevLett.91.097902} {\bibfield
  {journal} {\bibinfo  {journal} {Phys. Rev. Lett.}\ }\textbf {\bibinfo
  {volume} {91}},\ \bibinfo {pages} {097902} (\bibinfo {year}
  {2003})}\BibitemShut {NoStop}%
\bibitem [{\citenamefont {Lombardi}\ \emph {et~al.}(2002)\citenamefont
  {Lombardi}, \citenamefont {Sciarrino}, \citenamefont {Popescu},\ and\
  \citenamefont {De~Martini}}]{lombardi02}%
  \BibitemOpen
  \bibfield  {author} {\bibinfo {author} {\bibfnamefont {E.}~\bibnamefont
  {Lombardi}}, \bibinfo {author} {\bibfnamefont {F.}~\bibnamefont {Sciarrino}},
  \bibinfo {author} {\bibfnamefont {S.}~\bibnamefont {Popescu}}, \ and\
  \bibinfo {author} {\bibfnamefont {F.}~\bibnamefont {De~Martini}},\ }\href
  {\doibase 10.1103/PhysRevLett.88.070402} {\bibfield  {journal} {\bibinfo
  {journal} {Phys. Rev. Lett.}\ }\textbf {\bibinfo {volume} {88}},\ \bibinfo
  {pages} {070402} (\bibinfo {year} {2002})}\BibitemShut {NoStop}%
\bibitem [{\citenamefont {Bartlett}\ \emph {et~al.}(2007)\citenamefont
  {Bartlett}, \citenamefont {Rudolph},\ and\ \citenamefont
  {Spekkens}}]{bartlett07}%
  \BibitemOpen
  \bibfield  {author} {\bibinfo {author} {\bibfnamefont {S.~D.}\ \bibnamefont
  {Bartlett}}, \bibinfo {author} {\bibfnamefont {T.}~\bibnamefont {Rudolph}}, \
  and\ \bibinfo {author} {\bibfnamefont {R.~W.}\ \bibnamefont {Spekkens}},\
  }\href {\doibase 10.1103/RevModPhys.79.555} {\bibfield  {journal} {\bibinfo
  {journal} {Rev. Mod. Phys.}\ }\textbf {\bibinfo {volume} {79}},\ \bibinfo
  {pages} {555} (\bibinfo {year} {2007})}\BibitemShut {NoStop}%
\bibitem [{\citenamefont {Salart}\ \emph {et~al.}(2010)\citenamefont {Salart},
  \citenamefont {Landry}, \citenamefont {Sangouard}, \citenamefont {Gisin},
  \citenamefont {Herrmann}, \citenamefont {Sanguinetti}, \citenamefont {Simon},
  \citenamefont {Sohler}, \citenamefont {Thew}, \citenamefont {Thomas},\ and\
  \citenamefont {Zbinden}}]{salart10}%
  \BibitemOpen
  \bibfield  {author} {\bibinfo {author} {\bibfnamefont {D.}~\bibnamefont
  {Salart}}, \bibinfo {author} {\bibfnamefont {O.}~\bibnamefont {Landry}},
  \bibinfo {author} {\bibfnamefont {N.}~\bibnamefont {Sangouard}}, \bibinfo
  {author} {\bibfnamefont {N.}~\bibnamefont {Gisin}}, \bibinfo {author}
  {\bibfnamefont {H.}~\bibnamefont {Herrmann}}, \bibinfo {author}
  {\bibfnamefont {B.}~\bibnamefont {Sanguinetti}}, \bibinfo {author}
  {\bibfnamefont {C.}~\bibnamefont {Simon}}, \bibinfo {author} {\bibfnamefont
  {W.}~\bibnamefont {Sohler}}, \bibinfo {author} {\bibfnamefont {R.~T.}\
  \bibnamefont {Thew}}, \bibinfo {author} {\bibfnamefont {A.}~\bibnamefont
  {Thomas}}, \ and\ \bibinfo {author} {\bibfnamefont {H.}~\bibnamefont
  {Zbinden}},\ }\href {\doibase 10.1103/PhysRevLett.104.180504} {\bibfield
  {journal} {\bibinfo  {journal} {Phys. Rev. Lett.}\ }\textbf {\bibinfo
  {volume} {104}},\ \bibinfo {pages} {180504} (\bibinfo {year}
  {2010})}\BibitemShut {NoStop}%
\bibitem [{\citenamefont {Takeda}\ \emph {et~al.}(2015)\citenamefont {Takeda},
  \citenamefont {Fuwa}, \citenamefont {van Loock},\ and\ \citenamefont
  {Furusawa}}]{takeda15}%
  \BibitemOpen
  \bibfield  {author} {\bibinfo {author} {\bibfnamefont {S.}~\bibnamefont
  {Takeda}}, \bibinfo {author} {\bibfnamefont {M.}~\bibnamefont {Fuwa}},
  \bibinfo {author} {\bibfnamefont {P.}~\bibnamefont {van Loock}}, \ and\
  \bibinfo {author} {\bibfnamefont {A.}~\bibnamefont {Furusawa}},\ }\href
  {\doibase 10.1103/PhysRevLett.114.100501} {\bibfield  {journal} {\bibinfo
  {journal} {Phys. Rev. Lett.}\ }\textbf {\bibinfo {volume} {114}},\ \bibinfo
  {pages} {100501} (\bibinfo {year} {2015})}\BibitemShut {NoStop}%
\bibitem [{\citenamefont {Beenakker}\ \emph {et~al.}(2003)\citenamefont
  {Beenakker}, \citenamefont {Emary}, \citenamefont {Kindermann},\ and\
  \citenamefont {van Velsen}}]{beenakker03}%
  \BibitemOpen
  \bibfield  {author} {\bibinfo {author} {\bibfnamefont {C.~W.~J.}\
  \bibnamefont {Beenakker}}, \bibinfo {author} {\bibfnamefont {C.}~\bibnamefont
  {Emary}}, \bibinfo {author} {\bibfnamefont {M.}~\bibnamefont {Kindermann}}, \
  and\ \bibinfo {author} {\bibfnamefont {J.~L.}\ \bibnamefont {van Velsen}},\
  }\href {\doibase 10.1103/PhysRevLett.91.147901} {\bibfield  {journal}
  {\bibinfo  {journal} {Phys. Rev. Lett.}\ }\textbf {\bibinfo {volume} {91}},\
  \bibinfo {pages} {147901} (\bibinfo {year} {2003})}\BibitemShut {NoStop}%
\bibitem [{\citenamefont {Samuelsson}\ \emph {et~al.}(2004)\citenamefont
  {Samuelsson}, \citenamefont {Sukhorukov},\ and\ \citenamefont
  {B\"uttiker}}]{samuelsson04}%
  \BibitemOpen
  \bibfield  {author} {\bibinfo {author} {\bibfnamefont {P.}~\bibnamefont
  {Samuelsson}}, \bibinfo {author} {\bibfnamefont {E.~V.}\ \bibnamefont
  {Sukhorukov}}, \ and\ \bibinfo {author} {\bibfnamefont {M.}~\bibnamefont
  {B\"uttiker}},\ }\href {\doibase 10.1103/PhysRevLett.92.026805} {\bibfield
  {journal} {\bibinfo  {journal} {Phys. Rev. Lett.}\ }\textbf {\bibinfo
  {volume} {92}},\ \bibinfo {pages} {026805} (\bibinfo {year}
  {2004})}\BibitemShut {NoStop}%
\bibitem [{\citenamefont {Samuelsson}\ and\ \citenamefont
  {B\"uttiker}(2005)}]{samuelsson05}%
  \BibitemOpen
  \bibfield  {author} {\bibinfo {author} {\bibfnamefont {P.}~\bibnamefont
  {Samuelsson}}\ and\ \bibinfo {author} {\bibfnamefont {M.}~\bibnamefont
  {B\"uttiker}},\ }\href {\doibase 10.1103/PhysRevB.71.245317} {\bibfield
  {journal} {\bibinfo  {journal} {Phys. Rev. B}\ }\textbf {\bibinfo {volume}
  {71}},\ \bibinfo {pages} {245317} (\bibinfo {year} {2005})}\BibitemShut
  {NoStop}%
\bibitem [{\citenamefont {Sherkunov}\ \emph {et~al.}(2012)\citenamefont
  {Sherkunov}, \citenamefont {d'Ambrumenil}, \citenamefont {Samuelsson},\ and\
  \citenamefont {B\"uttiker}}]{sherkunov12}%
  \BibitemOpen
  \bibfield  {author} {\bibinfo {author} {\bibfnamefont {Y.}~\bibnamefont
  {Sherkunov}}, \bibinfo {author} {\bibfnamefont {N.}~\bibnamefont
  {d'Ambrumenil}}, \bibinfo {author} {\bibfnamefont {P.}~\bibnamefont
  {Samuelsson}}, \ and\ \bibinfo {author} {\bibfnamefont {M.}~\bibnamefont
  {B\"uttiker}},\ }\href {\doibase 10.1103/PhysRevB.85.081108} {\bibfield
  {journal} {\bibinfo  {journal} {Phys. Rev. B}\ }\textbf {\bibinfo {volume}
  {85}},\ \bibinfo {pages} {081108} (\bibinfo {year} {2012})}\BibitemShut
  {NoStop}%
\bibitem [{\citenamefont {Samuelsson}\ \emph {et~al.}(2003)\citenamefont
  {Samuelsson}, \citenamefont {Sukhorukov},\ and\ \citenamefont
  {B\"uttiker}}]{samuelsson03}%
  \BibitemOpen
  \bibfield  {author} {\bibinfo {author} {\bibfnamefont {P.}~\bibnamefont
  {Samuelsson}}, \bibinfo {author} {\bibfnamefont {E.~V.}\ \bibnamefont
  {Sukhorukov}}, \ and\ \bibinfo {author} {\bibfnamefont {M.}~\bibnamefont
  {B\"uttiker}},\ }\href {\doibase 10.1103/PhysRevLett.91.157002} {\bibfield
  {journal} {\bibinfo  {journal} {Phys. Rev. Lett.}\ }\textbf {\bibinfo
  {volume} {91}},\ \bibinfo {pages} {157002} (\bibinfo {year}
  {2003})}\BibitemShut {NoStop}%
\bibitem [{\citenamefont {Sim}\ and\ \citenamefont {Sukhorukov}(2006)}]{sim06}%
  \BibitemOpen
  \bibfield  {author} {\bibinfo {author} {\bibfnamefont {H.-S.}\ \bibnamefont
  {Sim}}\ and\ \bibinfo {author} {\bibfnamefont {E.~V.}\ \bibnamefont
  {Sukhorukov}},\ }\href {\doibase 10.1103/PhysRevLett.96.020407} {\bibfield
  {journal} {\bibinfo  {journal} {Phys. Rev. Lett.}\ }\textbf {\bibinfo
  {volume} {96}},\ \bibinfo {pages} {020407} (\bibinfo {year}
  {2006})}\BibitemShut {NoStop}%
\bibitem [{\citenamefont {Samuelsson}\ \emph {et~al.}(2009)\citenamefont
  {Samuelsson}, \citenamefont {Neder},\ and\ \citenamefont
  {B\"uttiker}}]{samuelsson09}%
  \BibitemOpen
  \bibfield  {author} {\bibinfo {author} {\bibfnamefont {P.}~\bibnamefont
  {Samuelsson}}, \bibinfo {author} {\bibfnamefont {I.}~\bibnamefont {Neder}}, \
  and\ \bibinfo {author} {\bibfnamefont {M.}~\bibnamefont {B\"uttiker}},\
  }\href {\doibase 10.1103/PhysRevLett.102.106804} {\bibfield  {journal}
  {\bibinfo  {journal} {Phys. Rev. Lett.}\ }\textbf {\bibinfo {volume} {102}},\
  \bibinfo {pages} {106804} (\bibinfo {year} {2009})}\BibitemShut {NoStop}%
\bibitem [{\citenamefont {Brunner}\ \emph {et~al.}(2014)\citenamefont
  {Brunner}, \citenamefont {Cavalcanti}, \citenamefont {Pironio}, \citenamefont
  {Scarani},\ and\ \citenamefont {Wehner}}]{brunner14}%
  \BibitemOpen
  \bibfield  {author} {\bibinfo {author} {\bibfnamefont {N.}~\bibnamefont
  {Brunner}}, \bibinfo {author} {\bibfnamefont {D.}~\bibnamefont {Cavalcanti}},
  \bibinfo {author} {\bibfnamefont {S.}~\bibnamefont {Pironio}}, \bibinfo
  {author} {\bibfnamefont {V.}~\bibnamefont {Scarani}}, \ and\ \bibinfo
  {author} {\bibfnamefont {S.}~\bibnamefont {Wehner}},\ }\href {\doibase
  10.1103/RevModPhys.86.419} {\bibfield  {journal} {\bibinfo  {journal} {Rev.
  Mod. Phys.}\ }\textbf {\bibinfo {volume} {86}},\ \bibinfo {pages} {419}
  (\bibinfo {year} {2014})}\BibitemShut {NoStop}%
\bibitem [{\citenamefont {Ji}\ \emph {et~al.}(2003)\citenamefont {Ji},
  \citenamefont {Chung}, \citenamefont {Sprinzak}, \citenamefont {Heiblum},
  \citenamefont {Mahalu},\ and\ \citenamefont {Shtrikman}}]{ji03}%
  \BibitemOpen
  \bibfield  {author} {\bibinfo {author} {\bibfnamefont {Y.}~\bibnamefont
  {Ji}}, \bibinfo {author} {\bibfnamefont {Y.}~\bibnamefont {Chung}}, \bibinfo
  {author} {\bibfnamefont {D.}~\bibnamefont {Sprinzak}}, \bibinfo {author}
  {\bibfnamefont {M.}~\bibnamefont {Heiblum}}, \bibinfo {author} {\bibfnamefont
  {D.}~\bibnamefont {Mahalu}}, \ and\ \bibinfo {author} {\bibfnamefont
  {H.}~\bibnamefont {Shtrikman}},\ }\href {\doibase 10.1038/nature01503}
  {\bibfield  {journal} {\bibinfo  {journal} {Nature}\ }\textbf {\bibinfo
  {volume} {422}},\ \bibinfo {pages} {415} (\bibinfo {year}
  {2003})}\BibitemShut {NoStop}%
\bibitem [{\citenamefont {Neder}\ \emph {et~al.}(2007)\citenamefont {Neder},
  \citenamefont {Ofek}, \citenamefont {Chung}, \citenamefont {Heiblum},
  \citenamefont {Mahalu},\ and\ \citenamefont {Umansky}}]{neder07}%
  \BibitemOpen
  \bibfield  {author} {\bibinfo {author} {\bibfnamefont {I.}~\bibnamefont
  {Neder}}, \bibinfo {author} {\bibfnamefont {N.}~\bibnamefont {Ofek}},
  \bibinfo {author} {\bibfnamefont {Y.}~\bibnamefont {Chung}}, \bibinfo
  {author} {\bibfnamefont {M.}~\bibnamefont {Heiblum}}, \bibinfo {author}
  {\bibfnamefont {D.}~\bibnamefont {Mahalu}}, \ and\ \bibinfo {author}
  {\bibfnamefont {V.}~\bibnamefont {Umansky}},\ }\href {\doibase
  10.1038/nature05955} {\bibfield  {journal} {\bibinfo  {journal} {Nature}\
  }\textbf {\bibinfo {volume} {448}},\ \bibinfo {pages} {333} (\bibinfo {year}
  {2007})}\BibitemShut {NoStop}%
\bibitem [{\citenamefont {Roulleau}\ \emph {et~al.}(2007)\citenamefont
  {Roulleau}, \citenamefont {Portier}, \citenamefont {Glattli}, \citenamefont
  {Roche}, \citenamefont {Cavanna}, \citenamefont {Faini}, \citenamefont
  {Gennser},\ and\ \citenamefont {Mailly}}]{roulleau07}%
  \BibitemOpen
  \bibfield  {author} {\bibinfo {author} {\bibfnamefont {P.}~\bibnamefont
  {Roulleau}}, \bibinfo {author} {\bibfnamefont {F.}~\bibnamefont {Portier}},
  \bibinfo {author} {\bibfnamefont {D.~C.}\ \bibnamefont {Glattli}}, \bibinfo
  {author} {\bibfnamefont {P.}~\bibnamefont {Roche}}, \bibinfo {author}
  {\bibfnamefont {A.}~\bibnamefont {Cavanna}}, \bibinfo {author} {\bibfnamefont
  {G.}~\bibnamefont {Faini}}, \bibinfo {author} {\bibfnamefont
  {U.}~\bibnamefont {Gennser}}, \ and\ \bibinfo {author} {\bibfnamefont
  {D.}~\bibnamefont {Mailly}},\ }\href {\doibase 10.1103/PhysRevB.76.161309}
  {\bibfield  {journal} {\bibinfo  {journal} {Phys. Rev. B}\ }\textbf {\bibinfo
  {volume} {76}},\ \bibinfo {pages} {161309} (\bibinfo {year}
  {2007})}\BibitemShut {NoStop}%
\bibitem [{\citenamefont {Litvin}\ \emph {et~al.}(2008)\citenamefont {Litvin},
  \citenamefont {Helzel}, \citenamefont {Tranitz}, \citenamefont
  {Wegscheider},\ and\ \citenamefont {Strunk}}]{litvin08}%
  \BibitemOpen
  \bibfield  {author} {\bibinfo {author} {\bibfnamefont {L.~V.}\ \bibnamefont
  {Litvin}}, \bibinfo {author} {\bibfnamefont {A.}~\bibnamefont {Helzel}},
  \bibinfo {author} {\bibfnamefont {H.-P.}\ \bibnamefont {Tranitz}}, \bibinfo
  {author} {\bibfnamefont {W.}~\bibnamefont {Wegscheider}}, \ and\ \bibinfo
  {author} {\bibfnamefont {C.}~\bibnamefont {Strunk}},\ }\href {\doibase
  10.1103/PhysRevB.78.075303} {\bibfield  {journal} {\bibinfo  {journal} {Phys.
  Rev. B}\ }\textbf {\bibinfo {volume} {78}},\ \bibinfo {pages} {075303}
  (\bibinfo {year} {2008})}\BibitemShut {NoStop}%
\bibitem [{\citenamefont {Huynh}\ \emph {et~al.}(2012)\citenamefont {Huynh},
  \citenamefont {Portier}, \citenamefont {le~Sueur}, \citenamefont {Faini},
  \citenamefont {Gennser}, \citenamefont {Mailly}, \citenamefont {Pierre},
  \citenamefont {Wegscheider},\ and\ \citenamefont {Roche}}]{huynh12}%
  \BibitemOpen
  \bibfield  {author} {\bibinfo {author} {\bibfnamefont {P.-A.}\ \bibnamefont
  {Huynh}}, \bibinfo {author} {\bibfnamefont {F.}~\bibnamefont {Portier}},
  \bibinfo {author} {\bibfnamefont {H.}~\bibnamefont {le~Sueur}}, \bibinfo
  {author} {\bibfnamefont {G.}~\bibnamefont {Faini}}, \bibinfo {author}
  {\bibfnamefont {U.}~\bibnamefont {Gennser}}, \bibinfo {author} {\bibfnamefont
  {D.}~\bibnamefont {Mailly}}, \bibinfo {author} {\bibfnamefont
  {F.}~\bibnamefont {Pierre}}, \bibinfo {author} {\bibfnamefont
  {W.}~\bibnamefont {Wegscheider}}, \ and\ \bibinfo {author} {\bibfnamefont
  {P.}~\bibnamefont {Roche}},\ }\href {\doibase 10.1103/PhysRevLett.108.256802}
  {\bibfield  {journal} {\bibinfo  {journal} {Phys. Rev. Lett.}\ }\textbf
  {\bibinfo {volume} {108}},\ \bibinfo {pages} {256802} (\bibinfo {year}
  {2012})}\BibitemShut {NoStop}%
\bibitem [{\citenamefont {Sherkunov}\ \emph {et~al.}(2009)\citenamefont
  {Sherkunov}, \citenamefont {Zhang}, \citenamefont {d'Ambrumenil},\ and\
  \citenamefont {Muzykantskii}}]{sherkunov09}%
  \BibitemOpen
  \bibfield  {author} {\bibinfo {author} {\bibfnamefont {Y.}~\bibnamefont
  {Sherkunov}}, \bibinfo {author} {\bibfnamefont {J.}~\bibnamefont {Zhang}},
  \bibinfo {author} {\bibfnamefont {N.}~\bibnamefont {d'Ambrumenil}}, \ and\
  \bibinfo {author} {\bibfnamefont {B.}~\bibnamefont {Muzykantskii}},\ }\href
  {\doibase 10.1103/PhysRevB.80.041313} {\bibfield  {journal} {\bibinfo
  {journal} {Phys. Rev. B}\ }\textbf {\bibinfo {volume} {80}},\ \bibinfo
  {pages} {041313(R)} (\bibinfo {year} {2009})}\BibitemShut {NoStop}%
\bibitem [{\citenamefont {Zhang}\ \emph {et~al.}(2009)\citenamefont {Zhang},
  \citenamefont {Sherkunov}, \citenamefont {d'Ambrumenil},\ and\ \citenamefont
  {Muzykantskii}}]{zhang09}%
  \BibitemOpen
  \bibfield  {author} {\bibinfo {author} {\bibfnamefont {J.}~\bibnamefont
  {Zhang}}, \bibinfo {author} {\bibfnamefont {Y.}~\bibnamefont {Sherkunov}},
  \bibinfo {author} {\bibfnamefont {N.}~\bibnamefont {d'Ambrumenil}}, \ and\
  \bibinfo {author} {\bibfnamefont {B.}~\bibnamefont {Muzykantskii}},\ }\href
  {\doibase 10.1103/PhysRevB.80.245308} {\bibfield  {journal} {\bibinfo
  {journal} {Phys. Rev. B}\ }\textbf {\bibinfo {volume} {80}},\ \bibinfo
  {pages} {245308} (\bibinfo {year} {2009})}\BibitemShut {NoStop}%
\bibitem [{\citenamefont {Beenakker}(2014)}]{beenakker14}%
  \BibitemOpen
  \bibfield  {author} {\bibinfo {author} {\bibfnamefont {C.~W.~J.}\
  \bibnamefont {Beenakker}},\ }\href {\doibase 10.1103/PhysRevLett.112.070604}
  {\bibfield  {journal} {\bibinfo  {journal} {Phys. Rev. Lett.}\ }\textbf
  {\bibinfo {volume} {112}},\ \bibinfo {pages} {070604} (\bibinfo {year}
  {2014})}\BibitemShut {NoStop}%
\bibitem [{\citenamefont {Burkard}\ and\ \citenamefont
  {Loss}(2003)}]{burkard03}%
  \BibitemOpen
  \bibfield  {author} {\bibinfo {author} {\bibfnamefont {G.}~\bibnamefont
  {Burkard}}\ and\ \bibinfo {author} {\bibfnamefont {D.}~\bibnamefont {Loss}},\
  }\href {\doibase 10.1103/PhysRevLett.91.087903} {\bibfield  {journal}
  {\bibinfo  {journal} {Phys. Rev. Lett.}\ }\textbf {\bibinfo {volume} {91}},\
  \bibinfo {pages} {087903} (\bibinfo {year} {2003})}\BibitemShut {NoStop}%
\bibitem [{\citenamefont {Giovannetti}\ \emph {et~al.}(2006)\citenamefont
  {Giovannetti}, \citenamefont {Frustaglia}, \citenamefont {Taddei},\ and\
  \citenamefont {Fazio}}]{giovanetti06}%
  \BibitemOpen
  \bibfield  {author} {\bibinfo {author} {\bibfnamefont {V.}~\bibnamefont
  {Giovannetti}}, \bibinfo {author} {\bibfnamefont {D.}~\bibnamefont
  {Frustaglia}}, \bibinfo {author} {\bibfnamefont {F.}~\bibnamefont {Taddei}},
  \ and\ \bibinfo {author} {\bibfnamefont {R.}~\bibnamefont {Fazio}},\ }\href
  {\doibase 10.1103/PhysRevB.74.115315} {\bibfield  {journal} {\bibinfo
  {journal} {Phys. Rev. B}\ }\textbf {\bibinfo {volume} {74}},\ \bibinfo
  {pages} {115315} (\bibinfo {year} {2006})}\BibitemShut {NoStop}%
\bibitem [{\citenamefont {Amico}\ \emph {et~al.}(2008)\citenamefont {Amico},
  \citenamefont {Fazio}, \citenamefont {Osterloh},\ and\ \citenamefont
  {Vedral}}]{amico08}%
  \BibitemOpen
  \bibfield  {author} {\bibinfo {author} {\bibfnamefont {L.}~\bibnamefont
  {Amico}}, \bibinfo {author} {\bibfnamefont {R.}~\bibnamefont {Fazio}},
  \bibinfo {author} {\bibfnamefont {A.}~\bibnamefont {Osterloh}}, \ and\
  \bibinfo {author} {\bibfnamefont {V.}~\bibnamefont {Vedral}},\ }\href
  {\doibase 10.1103/RevModPhys.80.517} {\bibfield  {journal} {\bibinfo
  {journal} {Rev. Mod. Phys.}\ }\textbf {\bibinfo {volume} {80}},\ \bibinfo
  {pages} {517} (\bibinfo {year} {2008})}\BibitemShut {NoStop}%
\bibitem [{Note1()}]{Note1}%
  \BibitemOpen
  \bibinfo {note} {See supplemental material at \protect \dots}\BibitemShut
  {NoStop}%
\bibitem [{\citenamefont {Horodecki}\ \emph {et~al.}(2009)\citenamefont
  {Horodecki}, \citenamefont {Horodecki}, \citenamefont {Horodecki},\ and\
  \citenamefont {Horodecki}}]{horodecki09}%
  \BibitemOpen
  \bibfield  {author} {\bibinfo {author} {\bibfnamefont {R.}~\bibnamefont
  {Horodecki}}, \bibinfo {author} {\bibfnamefont {P.}~\bibnamefont
  {Horodecki}}, \bibinfo {author} {\bibfnamefont {M.}~\bibnamefont
  {Horodecki}}, \ and\ \bibinfo {author} {\bibfnamefont {K.}~\bibnamefont
  {Horodecki}},\ }\href {\doibase 10.1103/RevModPhys.81.865} {\bibfield
  {journal} {\bibinfo  {journal} {Rev. Mod. Phys.}\ }\textbf {\bibinfo {volume}
  {81}},\ \bibinfo {pages} {865} (\bibinfo {year} {2009})}\BibitemShut
  {NoStop}%
\bibitem [{\citenamefont {Moskalets}\ and\ \citenamefont
  {B\"uttiker}(2002)}]{moskalets02}%
  \BibitemOpen
  \bibfield  {author} {\bibinfo {author} {\bibfnamefont {M.}~\bibnamefont
  {Moskalets}}\ and\ \bibinfo {author} {\bibfnamefont {M.}~\bibnamefont
  {B\"uttiker}},\ }\href {\doibase 10.1103/PhysRevB.66.205320} {\bibfield
  {journal} {\bibinfo  {journal} {Phys. Rev. B}\ }\textbf {\bibinfo {volume}
  {66}},\ \bibinfo {pages} {205320} (\bibinfo {year} {2002})}\BibitemShut
  {NoStop}%
\bibitem [{\citenamefont {Moskalets}(2011)}]{moskaletsbook}%
  \BibitemOpen
  \bibfield  {author} {\bibinfo {author} {\bibfnamefont {M.}~\bibnamefont
  {Moskalets}},\ }\href@noop {} {\emph {\bibinfo {title} {Scattering Matrix
  Approach to Non-Stationary Quantum Transport}}}\ (\bibinfo  {publisher}
  {Imperial College Press (London)},\ \bibinfo {year} {2011})\BibitemShut {NoStop}%
\bibitem [{\citenamefont {Fertig}\ and\ \citenamefont
  {Halperin}(1987)}]{fertig87}%
  \BibitemOpen
  \bibfield  {author} {\bibinfo {author} {\bibfnamefont {H.~A.}\ \bibnamefont
  {Fertig}}\ and\ \bibinfo {author} {\bibfnamefont {B.~I.}\ \bibnamefont
  {Halperin}},\ }\href {\doibase 10.1103/PhysRevB.36.7969} {\bibfield
  {journal} {\bibinfo  {journal} {Phys. Rev. B}\ }\textbf {\bibinfo {volume}
  {36}},\ \bibinfo {pages} {7969} (\bibinfo {year} {1987})}\BibitemShut
  {NoStop}%
\bibitem [{\citenamefont {B\"uttiker}(1990)}]{buttiker90}%
  \BibitemOpen
  \bibfield  {author} {\bibinfo {author} {\bibfnamefont {M.}~\bibnamefont
  {B\"uttiker}},\ }\href {\doibase 10.1103/PhysRevB.41.7906} {\bibfield
  {journal} {\bibinfo  {journal} {Phys. Rev. B}\ }\textbf {\bibinfo {volume}
  {41}},\ \bibinfo {pages} {7906} (\bibinfo {year} {1990})}\BibitemShut
  {NoStop}%
\bibitem [{\citenamefont {Hofer}\ and\ \citenamefont {Flindt}(2014)}]{hofer14}%
  \BibitemOpen
  \bibfield  {author} {\bibinfo {author} {\bibfnamefont {P.~P.}\ \bibnamefont
  {Hofer}}\ and\ \bibinfo {author} {\bibfnamefont {C.}~\bibnamefont {Flindt}},\
  }\href {\doibase 10.1103/PhysRevB.90.235416} {\bibfield  {journal} {\bibinfo
  {journal} {Phys. Rev. B}\ }\textbf {\bibinfo {volume} {90}},\ \bibinfo
  {pages} {235416} (\bibinfo {year} {2014})}\BibitemShut {NoStop}%
\bibitem [{\citenamefont {Klich}\ and\ \citenamefont
  {Levitov}(2009)}]{klich09}%
  \BibitemOpen
  \bibfield  {author} {\bibinfo {author} {\bibfnamefont {I.}~\bibnamefont
  {Klich}}\ and\ \bibinfo {author} {\bibfnamefont {L.~S.}\ \bibnamefont
  {Levitov}},\ }\href {\doibase 10.1103/PhysRevLett.102.100502} {\bibfield
  {journal} {\bibinfo  {journal} {Phys. Rev. Lett.}\ }\textbf {\bibinfo
  {volume} {102}},\ \bibinfo {pages} {100502} (\bibinfo {year}
  {2009})}\BibitemShut {NoStop}%
\bibitem [{\citenamefont {Song}\ \emph {et~al.}(2011)\citenamefont {Song},
  \citenamefont {Flindt}, \citenamefont {Rachel}, \citenamefont {Klich},\ and\
  \citenamefont {Le~Hur}}]{song11}%
  \BibitemOpen
  \bibfield  {author} {\bibinfo {author} {\bibfnamefont {H.~F.}\ \bibnamefont
  {Song}}, \bibinfo {author} {\bibfnamefont {C.}~\bibnamefont {Flindt}},
  \bibinfo {author} {\bibfnamefont {S.}~\bibnamefont {Rachel}}, \bibinfo
  {author} {\bibfnamefont {I.}~\bibnamefont {Klich}}, \ and\ \bibinfo {author}
  {\bibfnamefont {K.}~\bibnamefont {Le~Hur}},\ }\href {\doibase
  10.1103/PhysRevB.83.161408} {\bibfield  {journal} {\bibinfo  {journal} {Phys.
  Rev. B}\ }\textbf {\bibinfo {volume} {83}},\ \bibinfo {pages} {161408}
  (\bibinfo {year} {2011})}\BibitemShut {NoStop}%
\bibitem [{\citenamefont {Song}\ \emph {et~al.}(2012)\citenamefont {Song},
  \citenamefont {Rachel}, \citenamefont {Flindt}, \citenamefont {Klich},
  \citenamefont {Laflorencie},\ and\ \citenamefont {Le~Hur}}]{song12}%
  \BibitemOpen
  \bibfield  {author} {\bibinfo {author} {\bibfnamefont {H.~F.}\ \bibnamefont
  {Song}}, \bibinfo {author} {\bibfnamefont {S.}~\bibnamefont {Rachel}},
  \bibinfo {author} {\bibfnamefont {C.}~\bibnamefont {Flindt}}, \bibinfo
  {author} {\bibfnamefont {I.}~\bibnamefont {Klich}}, \bibinfo {author}
  {\bibfnamefont {N.}~\bibnamefont {Laflorencie}}, \ and\ \bibinfo {author}
  {\bibfnamefont {K.}~\bibnamefont {Le~Hur}},\ }\href {\doibase
  10.1103/PhysRevB.85.035409} {\bibfield  {journal} {\bibinfo  {journal} {Phys.
  Rev. B}\ }\textbf {\bibinfo {volume} {85}},\ \bibinfo {pages} {035409}
  (\bibinfo {year} {2012})}\BibitemShut {NoStop}%
\bibitem [{\citenamefont {Thomas}\ and\ \citenamefont
  {Flindt}(2015)}]{thomas15}%
  \BibitemOpen
  \bibfield  {author} {\bibinfo {author} {\bibfnamefont {K.~H.}\ \bibnamefont
  {Thomas}}\ and\ \bibinfo {author} {\bibfnamefont {C.}~\bibnamefont
  {Flindt}},\ }\href {\doibase 10.1103/PhysRevB.91.125406} {\bibfield
  {journal} {\bibinfo  {journal} {Phys. Rev. B}\ }\textbf {\bibinfo {volume}
  {91}},\ \bibinfo {pages} {125406} (\bibinfo {year} {2015})}\BibitemShut
  {NoStop}%
\end{thebibliography}
\end{document}